      [1999/12/01 v1.4c Il Nuovo Cimento]
\documentclass{cimento}
\usepackage{graphicx}

\def\simlt{\lower.5ex\hbox{$\; \buildrel < \over \sim \;$}}
\def\simgt{\lower.5ex\hbox{$\; \buildrel > \over \sim \;$}}
\def\simpt{\lower.5ex\hbox{$\; \buildrel \propto \over \sim \;$}}
\def\msun{\mbox{ M}_\odot}

\title{Gravitational Lensing of Pregalactic 21 cm Radiation}
\author{R.~Benton~Metcalf\from{ins:x}\ETC}
\instlist{\inst{ins:x}  Max Plank Institut f\"ur Astrophysics,  
Karl-Schwarzchild-Str. 1, 85741 Garching, Germany}
\begin{document}

\maketitle

\begin{abstract}
 Low-frequency radio observations of neutral hydrogen during and
 before the epoch of cosmic reionization will provide hundreds of
 quasi-independent source planes, each of precisely known redshift, if
 a resolution of $\sim 1$ arcminutes or better can be attained.  These
 planes can be used to reconstruct the projected mass distribution of
 foreground material. 
A wide-area survey of 21 cm lensing would provide very sensitive
constraints on cosmological parameters, in particular on dark
energy. These are up to 20 times tighter than the constraints
obtainable from comparably sized, very deep surveys of galaxy lensing
although the best constraints come from combining data of the two
types. Any radio telescope capable of mapping the 21cm brightness
temperature with good frequency resolution ($\sim $ 0.05 MHz) over a
band of width $\simgt$~10 MHz should be able to make mass maps of high
quality.   If the reionization epoch is at $z\simlt 9$ very large amounts of cosmological
information will be accessible.  The planned Square Kilometer Array (SKA) should be capable of
mapping the mass with a resolution of a few arcminutes depending on
the reionization history of the universe and how successfully
foreground sources can be subtracted.  The Low-Frequency Array (LOFAR) will
be able to measure an accurate matter power spectrum if the same
conditions are met. 
\end{abstract}

\section{Introduction}

The pregalactic 21 cm radiation originates from a period when most of
the hydrogen in the universe was neutral before most of the galaxies
formed and emitted ionizing radiation.  This is often called the dark
ages.  At $z\sim 1100$  the universe cooled to the extent where nearly all
of the hydrogen became neutral. We know this from the spectrum of the
CMB.  Some time between $z=6$ and $z=30$ most of the hydrogen in the
universe reionized.   We know this from the lack of sufficient Lyman-$\alpha$ absorption in high
 redshift quasar spectra.  The 21 cm
line comes from the hyperfine splitting of the ground state of
hydrogen (coupling between the spin of the electron and proton).  It
is redshifted to several meters when it reaches us. 

There will be many statistically independent regions of 21~cm emission
at different redshifts (or frequencies) at a single position on the
sky.  Gravitational lensing will distort the images of these
independent sources in a coherent way.  By ``stacking'' up the 21~cm
maps in different frequencies the lensing distortion can be separated
from the intrinsic structure in the 21~cm emission
\cite{ZandZ2006,metcalf&white2006}.

There are several telescopes being planned and built to observe the
21~cm radiation from the epoch of reionization.   Among these are the 21
Centimeter Array (21CMA, formerly known as
PAST)\footnote{http://21cma.bao.ac.cn/}, the Mileura Widefield Array
(MWA) Low Frequency Demonstrator
(LFD)\footnote{ttp://www.haystack.mit.edu/ast/arrays/mwa/}, the core 
array of LOFAR (Low Frequency Array)\footnote{www.lofar.org} and the 
core of SKA (Square Kilometer Array)\footnote{www.skatelescope.org/}.  
Only LOFAR and SKA will have sufficient collecting area to be relevant
for gravitational lensing observations as presently conceived.

\section{Imaging Dark Matter}

When low-frequency radio telescopes become sufficiently powerful to
map the signal from high-redshift 21~cm emission/absorption within
$\sim $10 or more statistically independent bands, the noise in the
lensing maps of foreground mass will be limited by the number of
frequency bands and not by the noise in the temperature maps
themselves.  This noise level is called the irreducible 
level because it depends only on the statistics of the source and the
frequency range of the telescope.  The irreducible noise level is low
enough that a high fidelity map of the foreground mass distribution
could be made.
At SKA-like resolution objects with virial masses as small as $2\times10^{13}\msun$
would be clearly visible.  The 21~cm emission is intrinsically
correlated to a relevant degree for frequency separations below $\sim $0.05 MHz
on 1 arcmin scales so decreasing the bandwidth below this level
produces no improvement in the lensing noise
\cite{metcalf&white2006}.  Simulated surface density maps have been
made by ray-tracing through the Millennium simulation \cite{HMW07}.

In the near term the telescopes will not reach the irreducible noise
limit because of noise in the brightness temperature measurements.  
Figure~\ref{fig:Ckappa} shows the noise in multipole space for a
SKA-like telescope and a LOFAR-like telescope after 90 days of
observing (approximately 3 seasons).  Where the noise per mode is well below
the expected signal a high fidelity reconstruction can be made.  SKA should be
able to image the matter with a resolution of a few arc-minutes.  
Exceptional regions such as around large galaxy clusters would have
higher fidelity down to smaller scales.  A LOFAR-like telescope will probably not be
able to image typical fluctuations in the mass density.  
It has been assumed for figure~\ref{fig:Ckappa} that the universe
very rapidly reionized at $z=7$ and that the frequency range of the
observations goes down to 100~MHz or $z=13$ for SKA and
114~MHz or $z=11.5$ for LOFAR.  Reionization
could increase the signal or decrease it depending on how and when it
occurs. See \cite{metcalf&white2007} for details.

Increasing the collecting area of the core of SKA by a factor of two
would reduce the noise to close to the irreducible limit.  Increasing
the resolution of the telescope would reduce the irreducible noise
both because of the number of statistically independent redshift
slices increases and because the number of independent patches on the
sky increases.   If a resolution of $\sim $6 arcsec (fwhm) could be
achieved, every halo more massive than the Milky Way's ($\sim
10^{12}\msun$) would be clearly visible back to z$\sim $10!  Since
almost no halos more massive then $\sim 10^{10}\msun$ would have
formed by $z=10$ this would be a complete inventory of {\it all} halos above
$\sim 10^{12}\msun$. 

21 cm lensing surveys could be cross-correlated with any other survey
of foreground objects including galaxy lensing surveys to get
additional tomographic information.  Images of the mass distribution through
nearly the entire depth of the observable universe would be of
enormous value for the study of cosmology and galaxy formation, and a
very direct test for the existence of dark matter.


\section{Measuring Cosmological Parameters}

\begin{figure}
\includegraphics[width=10cm]{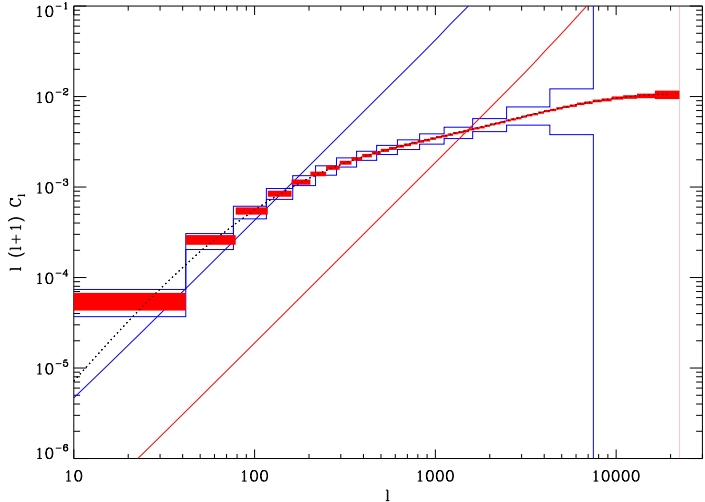}     
\caption{The estimated errors in the power spectrum of the projected
  density (convergence) 
fluctuations in the universe as a function of multipole number on the
sky.  The expected power spectrum is the dotted black curve.  The
almost straight curves are the noise per mode with blue for a
LOFAR-like telescope and red for a SKA-like telescope.  Where the expected signal
power spectrum is above the noise per mode ``typical'' fluctuations 
would be measurable, so a high fidelity map would be possible on these
scales.  This is also the range of scales where sample variance
would dominate the noise in the power spectrum estimate.
The blue outline marks the forecasted error in the power spectrum
estimate for a LOFAR-like telescope and the red filled boxes are for
an SKA-like telescope.  All calculations are done for 90 days of
observations (approximately 3 seasons) and 10\% of the sky surveyed.
See \cite{metcalf&white2007} for details.
}\label{fig:Ckappa}
\end{figure}

21 cm lensing is also capable of measuring the matter power spectrum and
its evolution.  Cross-correlating the convergence maps from 21 cm
lensing at different redshifts with each other and with galaxy lensing surveys
provides information on the  evolution of structure formation.   Any
cosmological parameters that affect the power spectrum and/or its
evolution can be probed in this way.  Dark energy is of particular
interest and 21 cm lensing would provide a unique probe of its
behavior at redshifts above $z\simeq 1$ as well as substantially constricting
the constraints on it below $z\simeq 1$. 

For estimating cosmological parameters the requirements on resolution
and frequency range are not as demanding as for imaging dark matter, but survey
area is of greater importance.  At the irreducible noise level the
uncertainties in the cosmological parameters are dominated by sample
variance if 10s of independent redshift slices are used.  Even for SKA,
sample variance should be the most important source of error for
multipole number $\ell \simlt 2000$ ($\ell \sim \pi/\theta =1.08\times 10^{4}$ arcmin /
$\theta$ ). This can be seen in figure~\ref{fig:Ckappa}.  Where the
noise per mode is below the expected signal the sample variance will
dominate in the power spectrum measurement.  The noise in the power spectrum estimate on
these scales can only be reduced by increasing the area of the survey.
For comparison, an ambitious all-sky galaxy lensing survey would reach the
sample variance limit at an $\ell$  of a few hundred. For the proposed radio
interferometers the survey area depends primarily on information
storage and processing speed which should improve over time. Table~\ref{table:params}
shows some estimates of the constraints that would be possible for a SKA-like
survey including modes up to $\ell = 1\times 10^4$ combined
with an idealized galaxy lensing survey.

The power spectrum of the projected mass density or convergence could
also be measured.  Figure~\ref{fig:Ckappa} shows the forecasted error
bars for SKA and LOFAR-like telescopes.  Both telescopes should be
capable of measuring the power spectrum well and SKA very well.

\begin{table}
\caption{Estimated constraints on some cosmological
  parameters.  The parameters are in order: the
  present density of dark energy, the power spectrum shape parameter, the equation
  of state parameter at z=0 (pressure/energy density), the derivative
  of the equation of state parameter with respect to the expansion
  parameter, the normalization of the power spectrum and the
  logarithmic slope of the primordial power spectrum.  The cosmology
  is assumed to be flat ($\Omega_m + \Omega_\Lambda=1$.  All errors have
  been marginalized over the other parameters in this set.  The galaxy
  survey has 35 galaxies per arc minute and covers half the sky as
  is expected for the next generation of ground and space based surveys
  (LSST, PanSTARRS, DUNE, etc.).  DUNE's wide survey would have
  a limiting magnitude of $I\simeq 24.5$ and a resolution of 0.23''
  for example.  Tomographic information has been
  used.  No photo-z errors are included.  The 21 cm survey is for 10
  redshift slices between z=7 and 13 and covers half the sky.  All errors are
  proportional to one over the square root of the fraction of the sky covered.
    The errors in parentheses are for the case of constant $w$ ($w_a=0$). 
}
\begin{tabular}{c|ll}
 & galaxies survey & galaxies survey alone \\
 & with SKA-like 21~cm survey  & \\
\hline
$\Delta\Omega_\Lambda$   & $4\times 10^{-4}$  & $0.001$   \\
$\Delta\Gamma$        & $3\times 10^{-4}$  & 0.004   \\
$\Delta w_o$              & 0.008 (0.008) & 0.03 (0.02)   \\
$\Delta w_a$            & 0.07 & 0.6   \\
$\Delta A$              & $0.006\times A$ & $0.8\times A$  \\
$\Delta n_s$            & $7\times 10^{-4}$ & 0.008     
\end{tabular}
\label{table:params}
\end{table}

\section{Conclusions}

Much will depend on future instrument design and the as yet unknown
characteristics of the 21 cm absorption and emission, particularly
around the epoch of reionization.  If reionization happens
unexpectedly early there may not be a large enough range of redshift
within the observed frequency range.  Despite this, the planned
specifications for SKA might enable us to make high fidelity maps of
the matter distribution all the way back to $z\simeq 10$ and, if enough
area can be surveyed, very good statistical information would be
possible enabling very high precision measurements of cosmological
parameters.  Realistic upgrades to the collecting area and array size 
of the next generation of telescopes would greatly improve their
sensitivity to lensing. 

\acknowledgments 
The author would like to thank the organizers for an excellent
conference and for the generous award.

 \bibliographystyle{/Users/bmetcalf/TeX/astronat-1.6/PhysRev/prsty}
 \bibliography{/Users/bmetcalf/Work/mybib}

\begin{thebibliography}{1}

\bibitem{ZandZ2006}
O. {Zahn} and M. {Zaldarriaga}, \apj {\bf 653},  922  (2006).

\bibitem{metcalf&white2006}
R.~B. {Metcalf} and S.~D.~M. {White}, \mnras {\bf 381},  447  (2007).

\bibitem{HMW07}
S. {Hilbert}, R.~B. {Metcalf}, and S.~D.~M. {White}, \mnras {\bf 706},  1106
  (2007).

\bibitem{metcalf&white2007}
R.~B. {Metcalf} and S.~D.~M. {White},   (2007), in preparation.

\end{thebibliography}

\end{document}